\documentclass[aps,twocolumn,tightenlines,a4paper]{revtex4}
\usepackage{graphicx}

\begin{document}
\draft

\title{Relativistic velocity singularities}

\author{M. J. Gagen}

\affiliation{Physics Department, The University of Queensland, QLD
4072 Australia }

\email{gagen@physics.uq.edu.au}

\date{\today}

\begin{abstract}
Relativistic and non-relativistic fluid equations can exhibit
finite time singular solutions including density singularities
appearing in collapse or compression systems and gradient
singularities in shock waves.  However, only the non-relativistic
fluid equations have been shown to exhibit finite-time velocity
singularities, $v\rightarrow\infty$ in this regime. As this limit
violates the cosmic speed limit $v<c$, where $c$ is the speed of
light, it is unclear how velocity singularity source terms in the
non-relativistic equations are modified in the relativistic
equations. This paper uses a driven system to demonstrate that
singular points within the relativistic fluid equations can
generate finite time relativistic velocity singularities
$v\rightarrow c$.
\end{abstract}






\maketitle

\section{Introduction}

The relativistic and non-relativistic fluid equations must
necessarily provide identical descriptions of fluid motion in the
non-relativistic regime suggesting that density and gradient
singularities present in the non-relativistic regime must also be
appear in the relativistic equations.  In particular, density
singularities $\rho\rightarrow\infty$ feature in the relativistic
gravitationally collapse of a non-rotating zero pressure dust
cloud \cite{Weinberg_72}, as well as in non-relativistic
compressive systems such as shock tunnels where growing pressures
burst thick metal diaphragms \cite{Glass_94}, in the hand-held
Diamond-Anvil generating 2 million atmospheres of pressure
\cite{Jayaraman_83_65}, in cavitation generating energy densities
sufficient to pit steel \cite{Young_89}, in sonoluminescence
generating temperatures in excess of 100,000 degrees
\cite{Weninger_97_17,Barber_97_65,Lohse_97_13}, in explosive
compression of magnetic fields generating fields up to 1,000T
\cite{Hill_97_48,Hill_97_20} and in supernovae collapse
experiments where inertial confinement generates temperatures in
excess of 100 million degrees \cite{Remington_97_19}.  Of course,
in the non-relativistic regime fluid parameters do not literally
become singular---the approximations deriving the non-relativistic
equations break down long before that point.

Similarly, gradient singularities appear within both the
relativistic and non-relativistic fluid equations to generate
shock waves within relativistic jets
\cite{Duncan_94_11,Marti_95_10}, in non-relativistic supersonic
flows \cite{Glass_74,Zucrow_76,Glass_94,Warsi_93}, and
interestingly, within trombones \cite{Hirschberg_96_17}.

Singular points present within the fluid equations can potentially
drive not only density and gradient singular solutions, but also
finite time velocity singularities.  In the non-relativistic
regime, velocity singularities $v\rightarrow\infty$ have been
exhibited in fissioning fluid drops
\cite{Shi_94_21,Kadanoff_97_11,Cohen_99_11,Zhang_99_11},
thin-films evaporating or pinched to zero thickness
\cite{Goldstein_93_30,Eggers_00_19}, and curvature collapse and
jet eruption from fluid surfaces \cite{Zeff_00_40}.  These systems
typically feature a free fluid surface of microscopic scale in one
dimension and macroscopic scale in other dimensions.  This scale
disparity allows surface tension to dominate other forces in the
microscopic dimensions allowing self-similar solutions where
essentially, fluid is evacuated across some macroscopic distance
on infinitesimal timescales generating a singular velocity
solution. These systems can be considerably simplified and scaled
up in size by replacing the free fluid surface by a flat moving
boundary which forces fluid evacuation over macroscopic distances
on infinitesimal timescales to again generate singular velocities.
The validity of this approach was experimentally confirmed in
Ref. \cite{Gagen_99_12}.

More generally, it is possible that similar singular source terms
might support finite time current singularities in
magnetohydrodynamic models of magnetic reconnection
\cite{Kerr_99_11}, and that unexplained car tire noise might
result from nonlinearities present in such driven collapse systems
\cite{Gagen_99_79}.  Additionally, velocity singularities appear
at boundary layer separation points \cite{Warsi_93}, while
examinations of non-relativistic fluid singularities and the
hydrodynamic blow-up problem appear in
\cite{Pumir_92_15,Ng_96_15,Pelz_97_16,Pelz_97_49,Grauer_98_41,Greene_00_79,Temam_00,Pelz_01_29}.

To date, there has been no examination of the potential for
singular points within the relativistic equations to drive
``relativistic velocity singularities" $v\rightarrow c$, where
here, $c$ is the speed of light recently termed ``Einstein's
constant" \cite{Brecher_2000}. Existing schemes to accelerate
particles to near light speed do not exploit velocity
singularities and typically rely on density singularities where
essentially, ``fireballs" accelerate particles as when coalescing
neutron stars source gamma ray bursts \cite{Eichler_89_12},
asymmetric (or turbulent) supernovae collapse driving bullets of
core material through overlying stellar layers
\cite{Hughes_00_L1,Umeda_00_L8}, and high intensity table-top
lasers drive particle acceleration \cite{Gahn_02_98}.  Other
proposed mechanisms employ gradual particle acceleration over
macroscopic distances and include shock wave surfing mechanisms
generating cosmic ray particle energies up to $10^{20}$eV
\cite{Longair_94}, and magnetohydrodynamic acceleration of polar
jets from accretion disks \cite{Blandford_82_88,Meier_97_35}.

In contrast to these approaches, this paper examines whether
singular points within the relativistic fluid equations can be
exploited to generate near relativistic flows.  The intractable
nature of the problem requires excessive simplification and, as in
the non-relativistic case \cite{Gagen_99_12}, we exploit driven
boundaries to force the relativistic fluid evacuation over
macroscopic distances on infinitesimal timescales, and demonstrate
in the appropriate limit that singular points within the
relativistic fluid equations can drive fluid velocities to the
limit $v\rightarrow c$.

\begin{figure}[htbp]
\includegraphics[width=6cm,clip]{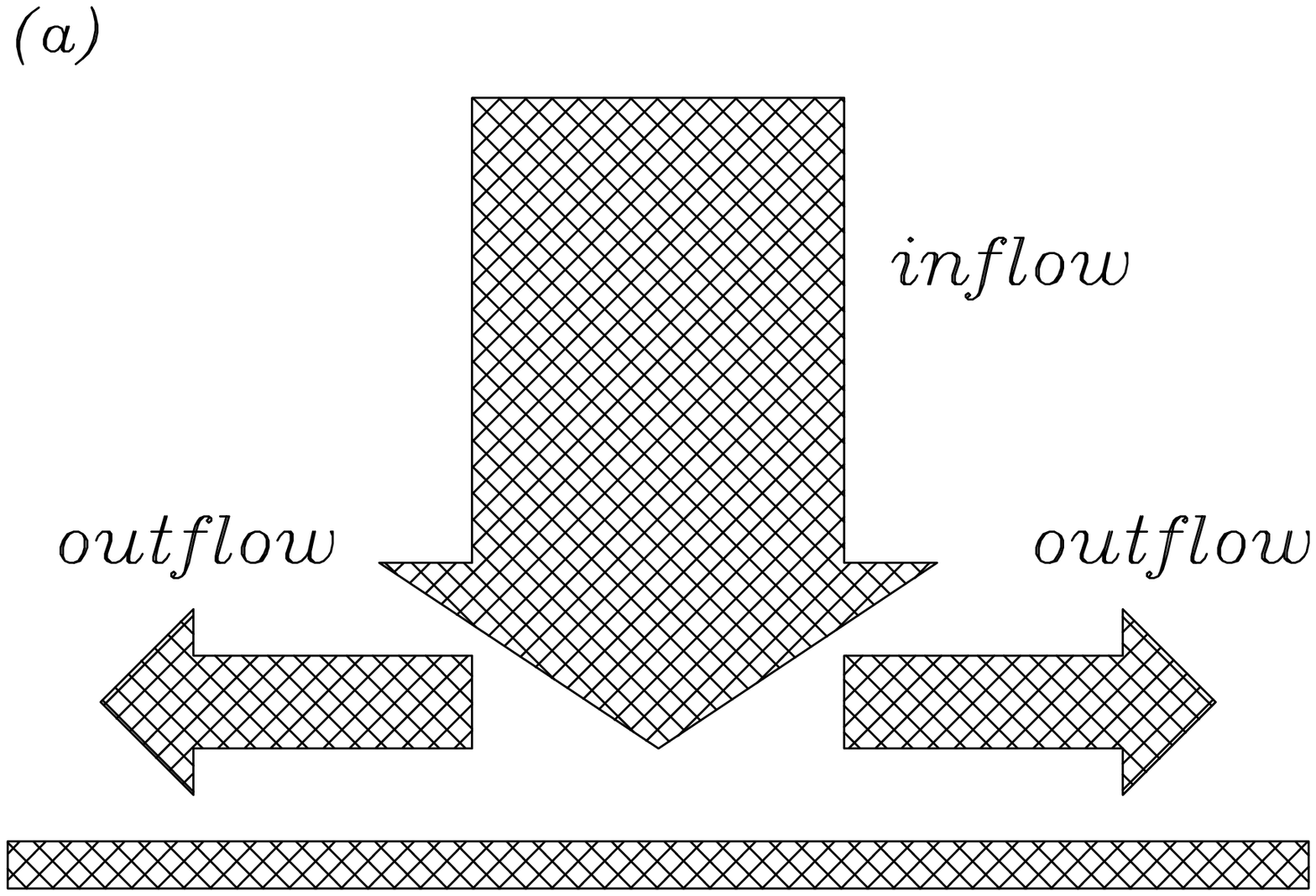}
\includegraphics[width=6cm,clip]{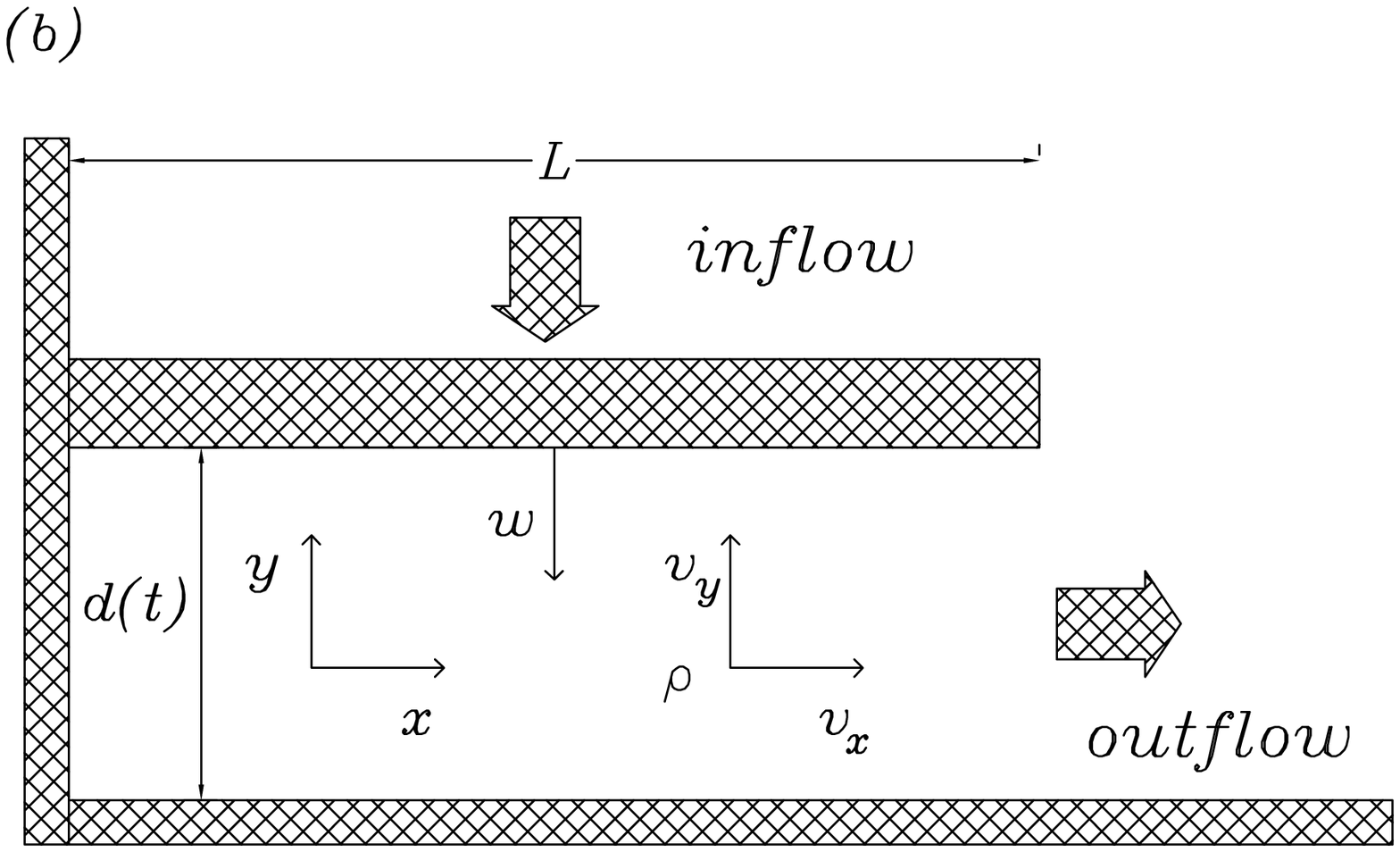}
\caption{ (a) A schematic of a near relativistic column of perfect
fluid impacting an incompressible surface, and (b) modelling the
column as a piston compressing underlying layers of gas. }
\label{f_squeeze}
\end{figure}

\section{Relativistic velocity singularities}

Our initial examination of relativistic velocity singularities
requires scaling up non-relativistic fluid evacuation systems to
relativistic speeds.  To this end, consider a perfect fluid column
travelling at near relativistic speeds hitting an incompressible
surface and generating an outflow as shown in Fig.
\ref{f_squeeze}(a).  The impacting fluid column acts as a
compressive piston as shown in Fig. \ref{f_squeeze}(b) so that as
the piston closes, entrapped fluid must evacuate over macroscopic
distances on ever shrinking timescales (assuming the fluid remains
relatively uncompressed.)  This crude model suffices to allow the
approximate solution of the fluid expulsion speed in horizontal
directions.

For simplicity we ignore gravity so fluid motion is entirely
governed by the relevant conservation equations, and consider a
flat space-time with metric
$\eta_{\alpha\beta}=\eta^{\alpha\beta}={\rm
diag}\left(-1,1,1,1\right)$ where primed coordinates $x'^{\alpha}
=\left(  t', x', y', z' \right)$ give the contravarient fluid
velocity vector as $U'^{\alpha}=\left(\gamma,\gamma{\bf
v}'\right)$ with ${\bf v}'=\left( v_{x'}, v_{y'}, v_{z'} \right)$
and $\gamma = 1 /\sqrt{1 - v'^2}$ in geometrized units setting the
speed of light to unity. (Greek indices run over values 0 --- 4
while Latin indices vary from 1---3.)

A perfect gas has particle flux and energy-momentum tensor
\begin{eqnarray}    \label{eq_primed_flux_energy}
   N'^{\alpha} & =& n U'^{\alpha} \nonumber \\
   T^{\alpha \beta} &=&  \eta^{\alpha \beta} p +
   \left(  p + \rho \right) U'^{\alpha} U'^{\beta}
\end{eqnarray}
where $n$, $p$ and $\rho$ are the momentarily comoving frame
particle number density, pressure and energy density respectively
\cite{Misner_73}. Conservation equations are ${N'^{\mu}}_{; \mu} =
0$ and ${T'^{\mu \nu}}_{; \mu} = 0$ giving
\begin{eqnarray}            \label{eq_prime}
 D' \left( \gamma n \right) & = & 0 \nonumber \\
 D' \big[ \gamma^2 (p + \rho) \big] - \partial_{t'} p &=& 0 \nonumber \\
 D' \big[ \gamma^2 (p + \rho) {\bf v'} \big] +
 {\bf \nabla'} p & = & 0,
\end{eqnarray}
with $D' = \partial_{t'} + {\bf \nabla' . v'}$ and ${\bf\nabla'}
=\left(\partial_{x'},\partial_{y'},\partial_{z'}\right)$. These
equations give respectively, particle number conservation, mass
continuity and momentum conservation.

The presence of a moving boundary (the descending piston)
complicates analysis and it is useful to use time dependent
computational space coordinates in which the descending piston is
rendered stationary.  If the piston is at height $d_0$ at time
$t=0$, then its subsequent height can be written $y'(t') = f(t')
d_0$ in terms of a monotonically decreasing function $f(t')$ with
$f(0)=1$.  Full closure occurs in the limit $f\rightarrow 0$, and
the piston's vertical velocity is $v_y (t') =\dot{f}(t') d_0$. The
transformation to unprimed computational space coordinates is then
\begin{equation}\label{eq_transform}
   x^{\alpha} = \left(
   \begin{array}{l}
   t  \\
   x  \\
   \chi \\
   z
   \end{array}
    \right)  \; = \;
    \left(
    \begin{array}{l}
   t'  \\
   x'  \\
   y'/f(t') \\
   z'
   \end{array}
    \right),
\end{equation}
so effectively, vertical compression in physical space $\Delta
y'\rightarrow 0$ corresponds to a constant separation in
computational coordinates, $\Delta\chi$ constant. The vertical
compression is fully captured in the function $f(t)=f(t')$, and
its effects can be assessed by tracking this function in governing
equations.

The new coordinates determine a new metric $g_{\mu \nu}$ and
proper time $d \tau^2=g_{\mu\nu} dx^{\mu} dx^{\nu}$
\begin{eqnarray}
d \tau^2 &=& \left[ \left( \dot{f} \chi \right)^2 -1 \right] dt^2
    + 2 f \dot{f} \chi dt \; d \chi \nonumber \\
    & &  + dx^2 + f^2 d \chi^2 + dz^2,
\end{eqnarray}
while the non-zero affine connections are
\begin{equation}
  \Gamma^{\chi}_{tt} = \frac{\ddot{f}}{f} \chi; \;\;\;\;\;
   \Gamma^{\chi}_{\chi t} = \Gamma^{\chi}_{t \chi}=
   \frac{\dot{f}}{f}.
\end{equation}
As required for a flat spacetime, the metric tensor $g_{\mu\nu}$
has three positive and one negative eigenvalue and the Ricci
tensor is zero everywhere. The transformed contravariant velocity
vector is
$U^{\alpha}=\left[\gamma,\gamma\left(v_x,v_{\chi},v_z\right)\right]$
with
\begin{equation}
  v_{\chi} = \frac{\left( v_{y'} -\dot{f}\chi \right)}{f}.
\end{equation}
Then, as required, the piston velocity is $v_{\chi}=0$ as
$v_{y'}=\dot{f}\chi (\equiv\dot{f} d_0)$.

In the new coordinates, the particle flux and energy-momentum
tensors become
\begin{eqnarray}    \label{eq_unprimed_flux_energy}
   N^{\alpha} & =& n U^{\alpha} \nonumber \\
   T^{\alpha \beta} &=&  g^{\alpha \beta} p +
   \left(  p + \rho \right) U^{\alpha} U^{\beta}
\end{eqnarray}
so the conservation equations ${N^{\mu}}_{;\mu} = 0$ and
${T^{\mu\nu}}_{;\mu} = 0$ give
\begin{eqnarray}            \label{eq_noprime}
 D \left(  f \gamma n \right) & = & 0 \nonumber \\
 D \left(  f \gamma^2 (p + \rho) \right) -  \partial_{t} (f p)
 & = & - \dot{f} \partial_{\chi} (\chi p)  \nonumber \\
 D \left(  f  \gamma^2 (p + \rho)  {\bf v} \right) + {\bf \nabla} \left( f p \right)
 & = & \left(
 \begin{array}{l}
  0  \\
  A  \\
  0
 \end{array}
 \right),
\end{eqnarray}
with
\begin{eqnarray}
  A &=& - \ddot{f} \chi T^{tt} - 2 \dot{f} T^{t\chi}
  - \partial_{t} \left( \dot{f} \chi p \right) \nonumber  \\
  & & - \partial_{\chi} \left[ f p  \left( \frac{1}{f^2} -
  \frac{\dot{f}^2 \chi^2}{f^2} - 1 \right) \right],
\end{eqnarray}
and $D = \partial_t + {\bf \nabla . v}$, ${\bf \nabla}
=\left(\partial_x, \partial_\chi, \partial_z \right)$, and
\begin{eqnarray}
  T^{tt}  & = & \gamma^2 (p+\rho)-p  \nonumber \\
  T^{t\chi} & = & \gamma^2 (p+\rho) v_{\chi}
       + \frac{\dot{f}}{f}  \chi p.
\end{eqnarray}
These equations give respectively, particle number conservation,
mass continuity and momentum conservation as can be seen by taking
the limit $f = 1$ and $\dot{f}=0$.

An understanding of the physics embodied in Eqs. (\ref{eq_prime})
and (\ref{eq_noprime}) can be obtained by considering solutions in
two exactly solvable physical limits. The first considers a
stationary piston with the fluid in equilibrium.  In this case,
Eq. (\ref{eq_prime}) is solved by setting fluid velocity ${\bf
v'}=0$, and number density, pressure and density everywhere
constant.  Eq. (\ref{eq_noprime}) reduces to Eq. (\ref{eq_prime})
in the case $f=1$, while for $f\leq 1$, Eq. (\ref{eq_noprime}) is
solved by setting ${\bf v}=(0,-\dot{f}\chi/f,0)$, and again,
uniform number density, pressure and density.  The second case
considers a collisionless relativistic dust with $p=0$ (as some
equation of state must be specified), and treats a closed piston
lacking horizontal flows $v_{x'}=v_{z'}=0$ with the piston
descending at a constant velocity ($\ddot{f}=0$).  The dust at
height $y$ has initial vertical velocity at time $t=0$ of
$v_{y'}=\dot{f} y$ and these dust elements preserve this velocity
for all times as pressure is zero. For later times $t$ at fixed
height $y$, the fluid velocity increases as $v_{y'}=\dot{f} y/f$.
Eq. (\ref{eq_prime}) is then solved by ${\bf v'}=(0,\dot{f}y/f,0)$
and everywhere uniformly increasing number density $\gamma
n(t)=c_1/f$ and fluid density $\gamma^2 \rho(t)=c_2/f$ for
constants $c_1$ and $c_2$. Conversely, the moving coordinate
system tracks the initial dust elements leading to a constant
velocity ${\bf v}=(0,0,0)$ which together with uniformly
increasing number density $\gamma n(t)=c_1/f$ and fluid density
$\gamma^2 \rho(t)=c_2/f$ solves Eq. (\ref{eq_noprime}).

We now turn to consider the role of the squeezing parameter $f(t)$
in Eq. (\ref{eq_noprime}) which is linked everywhere on the
left-hand-side (LHS) with $\gamma$. As is well known, the
singularity $\gamma\rightarrow\infty$ ensures that all fluid
velocities remain less than the speed of light, $v < c$. Then, the
countervailing linkage between this singularity and the zero
asymptote of the squeezing parameter $f\rightarrow 0$ suggests a
minimization of the effects of the combined terms $f\gamma$ and
$f\gamma^2$.

The physical meaning of the squeezing parameter $f$ can be seen in
more detail using the injection form of the relativistic fluid
equations
\begin{eqnarray}            \label{eq_injection}
 D \left(  \gamma n \right) & = & - \frac{\dot{f} \gamma n}{f} \nonumber \\
 D \left[  \gamma^2 (p + \rho) \right] - \partial_{t} p
 & = & - \frac{\dot{f}}{f} \partial_{\chi} (\chi p)  \nonumber \\
    && - \frac{\dot{f}}{f}
     \left( \gamma^2 (p+\rho) - p \right) \\
 D \left[ \gamma^2 (p + \rho)  {\bf v} \right]
          + {\bf \nabla} p
 & = & \left(
 \begin{array}{l}
  0  \\
  A/f  \\
  0
 \end{array}
 \right)
 - \frac{\dot{f}\gamma^2(p+\rho)}{f}
 \left(
 \begin{array}{l}
  v_x  \\
  v_{\chi}  \\
  v_z
 \end{array}
 \right) .\nonumber
\end{eqnarray}
The left hand sides (LHS) here are identical to those of Eq.
(\ref{eq_prime}) while the right hand sides (RHS) represent
sources injecting matter and momentum, including into the
horizontal directions $x$ and $z$. (Similar equations are used to
describe rockets employing real injection systems to generate high
speed expelled flows \cite{Zucrow_76}.) The injection sources are
dominated in the relativistic large $\gamma$ regime by terms
proportional to the ratio $\dot{f}/ f$ and either $\gamma$ or
$\gamma^2$. Then, as long as $\dot{f}\neq 0$, the respective
limits $\gamma\rightarrow\infty$ and $f\rightarrow 0$ reinforce
each other to inject (potentially) infinite mass and momentum in
finite time to drive a horizontal relativistic velocity
singularity.

A very approximate analysis of the velocity and energy of the
horizontal expelled jet can be performed but solutions are
suggestive only. Consider the case where again the moving
coordinate system approximately tracks the descending fluid so
$v_{\chi}\approx 0$, though now, the increasing pressure drives a
near relativistic horizontal velocity $v_x$ constituting an
expelled jet.  (Symmetry allows ignoring the $z$ direction.) Fluid
velocities are then $\left( v_x, 0, 0\right)$. In the large
$\gamma$ limit each of the above conservation equations gives the
time rate of change of the expulsion velocity as
\begin{eqnarray}            \label{eq_expelled_jet}
 \partial_t v_x &=&
 - \frac{{\bf \nabla . v}\gamma n+\gamma \dot{n}}{n\gamma^3 v_x}
 - \frac{\dot{f}}{f \gamma^2 v_x}  \nonumber \\
 \partial_t v_x &=& - \frac{{\bf \nabla . v}\gamma^2 R+\gamma^2 \dot{R}}{2 \gamma^4 R v_x}
 - \frac{\dot{f} (R-p) }{2 f R \gamma^2 v_x}  \nonumber \\
 \partial_t v_x &=&  - \frac{{\bf \nabla . v}\gamma^2 R v_x + \partial_x p}
 {\gamma^2 R(1+ 2 \gamma^2 v_x^2)}
 -\frac{\dot{f} v_x }{f (1+ 2 \gamma^2 v_x^2)}
\end{eqnarray}
where $R=\gamma^2 (p+\rho)$ and the top line results from the
conservation of particle number, the second line from the
conservation of mass and the third line stems from the
conservation of momentum in the $x$ direction.  The asymptotic
form of each of these equations in the large $\gamma$ limit is
approximately
\begin{equation}
 \partial_t v_x \approx O\left( \frac{1}{\gamma^2} \right) -
 \frac{\dot{f}}{f \gamma^2 v_x}
\end{equation}
and this form is used for further discussion. The first term on
the RHS is independent of $f$ and, in the absence of squeezing
($\dot{f} = 0$), necessarily goes to zero so that $\partial_t
v_x\rightarrow 0$ as $\gamma\rightarrow\infty$ ensuring $v<c$. The
second term reflects squeezing effects and is positive
($\dot{f}<0$) and the linkage between the limits $f\rightarrow 0$
and $\gamma\rightarrow\infty$ offsets the effect of $\gamma$. This
equation can be integrated by ignoring the small first term with
initial conditions $f(t_0)=1$ and $v_x(t_0)=v_0$ giving
\begin{equation}
  v^2_{x} (t)  =  1 - f^{2} \left(  1 - v^2_{0} \right),
\end{equation}
or equivalently
\begin{equation}         \label{eq_gamma}
  \gamma^2 (t) = \frac{\gamma_0^2}{f^{2}}.
\end{equation}
It is of interest to consider the non-physical mathematical limit
$f\rightarrow 0$ for some finite time $T>t_0$ representing full
closure so $f(T) = 0$.  This limit gives
\begin{equation}
  v_{x} (T) = 1,
\end{equation}
a light speed jet.  It is of course impossible for any material
fluid to reach light speed though squeezed system might be able to
exploit relativistic velocity singularity source terms to approach
this ideal using singular mass and momentum injection source terms
proportional to $-\dot{f}\gamma^2 / f$.

Particle energy within the expelled jet is then
\begin{equation}
  E \propto \gamma (t) \propto  \frac{1}{f} \rightarrow
  \infty,
\end{equation}
though this result is suggestive only.

\section{Non-relativistic velocity singularities}

For completeness, details of the non-relativistic velocity
singularity created by asymmetrically compressing a fluid are
given.  (See Ref. \cite{Gagen_99_12} for more details.) The
conservative form of the inviscid and dimensionless Euler
equations with zero conductivity are derived from the the
relativistic equations (\ref{eq_noprime}) after taking the small
velocity limit $\gamma\rightarrow 1$, $p\approx O(\rho v^2)$, and
$p + \rho\rightarrow\rho$, so now $\rho$ and $p$ become the usual
fluid density and pressure. The use of time-dependent coordinates
[Eq.~(\ref{eq_transform})] gives the Energy-Momentum tensor for a
perfect gas as
\begin{eqnarray}
   T^{00} &=& \rho  \nonumber  \\
   T^{0j} &=& \rho v^{j}   \nonumber \\
   T^{ij} &=& \rho v^{i} v^{j} +
          {\rm diag} \left( p, p / f^2 , p \right).
\end{eqnarray}
Here, the diagonal pressure terms in the vertical direction are
scaled by a factor $1/f^2$ over those in horizontal directions so
that the limit $f \rightarrow 0$ creates the large pressure
gradients driving the horizontal velocity singularity.
Conservation of the energy-momentum tensor ${T^{\mu \nu}}_{; \mu}
= 0$ then gives the modified Euler equations
\begin{equation}         \label{eq_Euler}
\partial_t U + \partial_x F + \partial_{\chi} G  =  0
\end{equation}
\[
\begin{array}{cl}
U = \left( \begin{array}{c}
              f \rho \\
              f  \rho v_x  \\
              f \rho v_y
             \end{array}\right)  &
F = \left (\begin{array}{c}
              f \rho v_x \\
              f \left( \rho v_x^2 + \rho^{\gamma}/\gamma \right) \\
              f \rho v_x v_y
             \end{array}\right) \\
G = \left( \begin{array}{c}
              f \rho v_{\chi} \\
              f \rho v_x v_{\chi}  \\
              f \rho v_y v_{\chi} +  \rho^{\gamma}/\gamma
             \end{array}\right )
\end{array}.
\]
Here $v_{\chi} =( v_y -\dot{f}\chi) / f$ and we show mass
continuity (top line) and momentum conservation in the $x$ and $y$
directions. (Symmetry constrains fluid motions to this plane.)
Adiabatic perfect gases with $p=\rho^{\gamma}$ ($\gamma=1.4$) are
considered and all variables are dimensionless with dimensioned
(primed) variables being given by $x' = x L$, $v' = v a_0$, $p' =
p p_0$, $\rho' =\rho\rho_0$ and $t' = t L / a_0$ with $L$ being
some convenient length parameter and $a_0^2=\gamma p_0 /\rho_0$
giving the local speed of sound.

Analytic solutions can be obtained in the low velocity limit
$v_x\ll 1$ (allowing $v_x^2 \approx 0$), and by assuming linear
squeezing $f(t)=1-t/T$ for some closure time $T$, with fluid
elements possessing velocity $v_y =\dot{f}\chi$ tracked by the
time dependent coordinates giving $v_{\chi}=0$. The analytic
solution for an incompressible fluid with $\rho = 1$ (to give the
maximum expulsion velocity) is then
\begin{equation}         \label{eq_vel}
  v_x (x,t) = - \frac{\dot{f} x}{f}
\end{equation}
displaying a velocity singularity as $f\rightarrow 0$. This
solution is strictly valid only while $v_x < 0.3$ where an
unconfined compressible fluid remains approximately uncompressed.
Experiments demonstrate that this solution provides a reasonable
approximation to the expulsion velocity into the supersonic regime
\cite{Gagen_99_12}.

\section{Conclusion}

This paper uses a preliminary and necessarily crude analysis to
establish that relativistic velocity singularities can appear in
relativistic hydrodynamic equations. Velocity singularities
provide such an elegant and simple means of sourcing near light
speed relativistic jets that it would be surprising if this
mechanism was not accessed in some astrophysical systems.

\vspace{-0.5cm}

\end{document}